\documentclass[
preprint,
preprintnumbers,
 amsmath,amssymb,
aip,
]{revtex4-2}

\usepackage{graphicx}
\usepackage{dcolumn}
\usepackage{bm}
\usepackage{xcolor}
\usepackage{hyperref}
\usepackage[mathlines]{lineno}

\begin{document}

\preprint{APS/123-QED}

\title{Anisotropic Response in Metamaterials with Elliptically Perforated Plates: Applications to Near-Field Radiative Heat Transfer} 

\author{J. E. P\'erez-Rodr\'{i}guez}
\affiliation{Instituto F\'{i}sica, Universidad Nacional Aut\'onoma de M\'exico, Cto. de la Investigación SN, CDMX 01000, Mexico.}

\author{R. Esquivel-Sirvent}
\email{raul@fisica.unam.mx}
\affiliation{Instituto F\'{i}sica, Universidad Nacional Aut\'onoma de M\'exico, Cto. de la Investigación
 SN, CDMX 01000, Mexico.}
 
\author{A. Camacho de la Rosa}
\affiliation{Centro de Investigaci\'on en Ciencias, Instituto de Investigaci\'on en Ciencias B\'asicas y Aplicadas, 62209 Cuernavaca, Morelos, M\'exico }

\date{\today}

\begin{abstract}
Metamaterials with tunable optical properties provide a versatile platform for controlling electromagnetic interactions at the nanoscale. This study explores the anisotropic thermal behavior of metamaterials composed of planar plates perforated with periodic arrays of cylinders possessing elliptical cross sections. In contrast to conventional circular perforations, elliptical geometries inherently break rotational symmetry, introducing anisotropy in the effective electromagnetic and thermal response of the structure. Using a fluctuation electrodynamics framework combined with full-wave numerical simulations, we quantify the near-field radiative heat transfer between such elliptically perforated plates as a function of ellipse orientation, aspect ratio, and separation distance. The results reveal that elliptical perforations enable enhanced spectral and directional control of evanescent mode coupling and surface polariton excitation, leading to significant modulation of the near-field heat flux. These findings highlight the potential of geometrically engineered anisotropy for advanced thermal management and energy conversion applications, and offer new design strategies for the development of thermally functional metamaterials operating in the near-field regime.
\end{abstract}

\maketitle


\section{Introduction}

Metamaterials are artificially structured media \cite{cui2010metamaterials} engineered to exhibit tailored responses such as  electromagnetic \cite{solymar2009waves,dubey2012metamaterials,sihvola2007metamaterials} or thermal fields and have enabled unprecedented control over wave propagation and energy transfer \cite{guo2012broadband,dyachenko2016controlling,coppens2017spatial}. One particularly promising application lies in the enhancement and manipulation of near-field radiative heat transfer (NFRHT), where the exchange of thermal energy between closely spaced bodies exceeds the blackbody limit due to tunneling of evanescent waves \cite{vanhove,HARGREAVES,Song}. The geometry of the constituent structures in metamaterials plays a pivotal role in shaping their effective properties and determining the efficiency of near-field interactions \cite{Effective,Francoeur15,gratings,PhysRevB.85.155418,Liu}.

In this work, we focus on metamaterials composed of periodic arrays of cylinders with elliptical cross sections, perforating a planar substrate. Compared to their circular counterparts \cite{biehs2011nanoscale}, elliptically shaped cylinders offer several key advantages. Most notably, the elliptical geometry introduces inherent anisotropy in the material response, which can be finely tuned by adjusting the aspect ratio and orientation of the ellipses. This  anisotropy enables directional control of thermal radiation and enhanced coupling \cite{fang2022anisotropic,mia2020exceptional} to anisotropic surface modes, both of which are crucial for optimizing NFRHT. These properties are especially valuable in applications such as thermal rectification \cite{otey2010thermal,li2021transforming}, and directional heat shields \cite{raza2016transformation}, where direction-sensitive control of heat is essential. 
As in Biehs et al. \cite{biehs2011nanoscale} we employ an effective medium approach (EMA). For other systems, EMA has proven useful in the calculation of NFRHT in systems such as multilayered metamaterials \cite{liu2014application}, rough surfaces \cite{xu2019near}, and composite materials \cite{perez2020revisiting}. 

In this work, we investigate how the anisotropic nature of elliptically perforated plates influences the near-field thermal exchange between parallel structures. The results reveal new pathways to engineer thermal metamaterials with tunable, directional heat transfer characteristics—extending the functional versatility of near-field thermal technologies beyond what is possible with isotropic (circular) geometries.

\section{Theory}
A homogeneous, isotropic slab characterized by a local dielectric function $\epsilon_h$ is considered. The material contains an array of cylindrical holes of radius $a$, each with a dielectric function $\epsilon_i$. In general, both dielectric functions are frequency dependent. The volume filling fraction $f$ is defined as the ratio of the volume occupied by the cylinders to the total volume. Using an effective medium approximation, the composite dielectric response can be approximated analytically. For circular cylindrical inclusions, closed-form expressions for the effective dielectric function are obtained as \cite{biehs2011nanoscale,choy2015effective}: 
\begin{equation}
\epsilon_{||}=\epsilon_h \frac{\epsilon_i(1+f)+\epsilon_h(1-f)}{\epsilon_i(1-f)+\epsilon_h(1+f)},
\end{equation}
and
\begin{equation}
\epsilon_{\bot}=\epsilon_h (1-f)+\epsilon_i f,
\end{equation}
where $\epsilon_{||}=\epsilon_z$ is the dielectric function parallel to the axis of the cylinders, and $\epsilon_{\bot}=\epsilon_{xx}=\epsilon_{yy}$ are the components parallel to the slab surface. Thus, the homogenization procedure yields an effective dielectric tensor:
\[
\bar{\bar{\epsilon}} = \mathrm{diag}(\epsilon_{xx},\epsilon_{yy},\epsilon_{zz}) = \mathrm{diag}(\epsilon_{\bot},\epsilon_{\bot},\epsilon_{||}).
\]

We now consider cylinders with elliptical cross-sections, as shown in Fig.~1(b), with major axis $a$ and minor axis $b$. The cross section of the cylinders will have an eccentricity $e=\sqrt{1-(b/a)^2}$. The effective dielectric permittivity for elliptical cylinders can be obtained using a Maxwell-Garnett-type approximation:
\begin{equation}
\frac{\epsilon_{j}-\epsilon_h}{\epsilon_h + L^{(j)} (\epsilon_{j}-\epsilon_h)} = 
f \frac{\epsilon_i-\epsilon_h}{\epsilon_h + L^{(j)} (\epsilon_i-\epsilon_h)},
\label{eq:MGgeneral}
\end{equation}
where $j = x, y, z$, and $L^{(j)}$ are the depolarization (or form) factors, given by
\begin{equation}
L^{(x)}(\sigma_{ab}) = \frac{ab}{a^2-b^2} \int_{\sigma_{ab}}^{\infty} \frac{ds}{s^2 \sqrt{s^2-1}},
\label{eq:EllipticCilynder1}
\end{equation}
and
\begin{equation}
L^{(y)}(\sigma_{ab}) = \frac{ab}{a^2-b^2} \int_{\sigma_{ab}}^{\infty} \frac{ds}{(s^{2}-1)^{3/2}},
\label{eq:EllipticCilynder2}
\end{equation}
where $x$ is along the major axis and $y$ along the minor axis. The lower limit of the integral is
\begin{equation}
\sigma_{ab} = \frac{a}{\sqrt{a^{2}-b^{2}}}.
\label{eq:Sigmaab}
\end{equation}

For very long and narrow cylinders, the depolarization factor along the axis is $L^{(z)} = 0$. These form factors satisfy the closure relation $L^{(x)} + L^{(y)} = 1$. For circular cross-sections ($a = b$), we have $L^{(x)} = L^{(y)} = 1/2$. For elliptical cylinders, the resulting effective dielectric tensor corresponds to a biaxial medium:
\begin{equation}
\bar{\bar{\epsilon}} = \mathrm{diag}(\epsilon_{xx}, \epsilon_{yy}, \epsilon_{zz}).
\label{tensordiel}
\end{equation}
The details of the derivation of the form factors are provided in the Supplementary Material section as well as the dependence of $L_x$ and $L_y$ with eccentricity in Fig.(S1).   

\begin{figure}[ht]
\centering
  \includegraphics[width=0.6\textwidth]{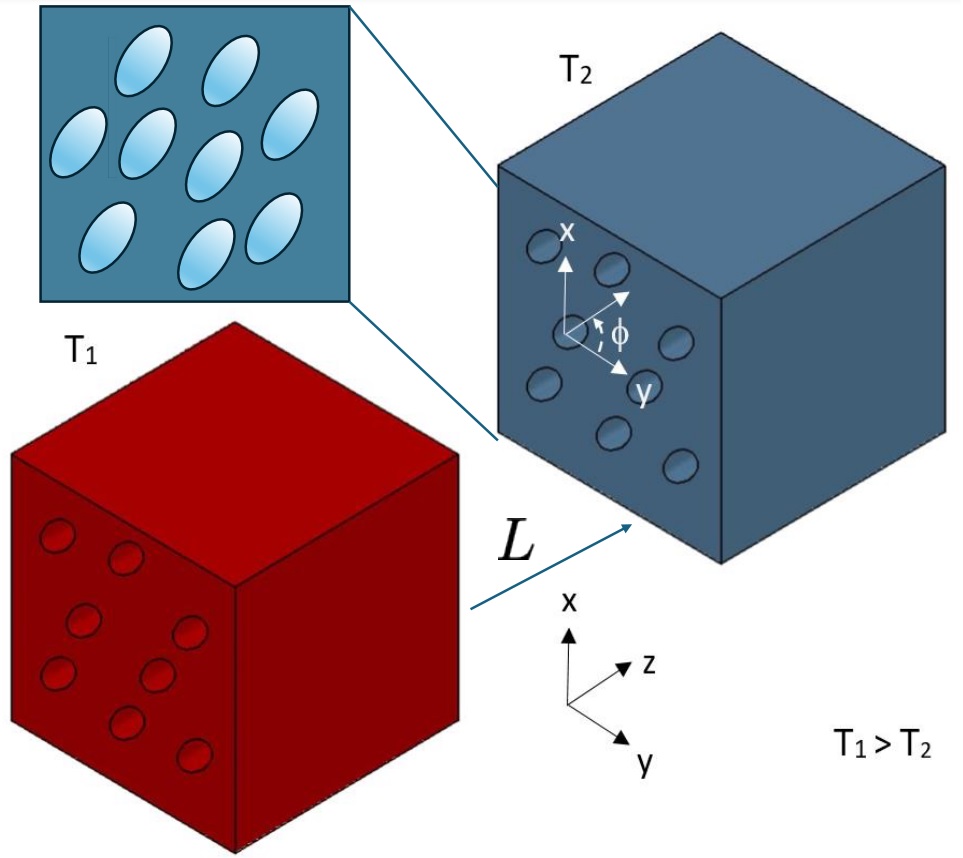}
  \caption{Schematics of the system under study, including a front view of the slabs. Two parallel slabs with parallel perforations that can be of circular cross section or elliptical cross section. The cylinders are randomly placed and occupy a volume fraction $f$. The angle $\phi$ is measured with respect the $y$ axis, as indicated.  }
  \label{Fig1}
 \end{figure}

Once the effective dielectric function of the system is obtained, we consider two identical slabs containing the cylindrical inclusions, separated by a gap as shown in Fig.~\ref{Fig1}. The theory for calculating NFRHT is well established within the framework of thermally excited electromagnetic fields \cite{Vinogradov} . Given the temperatures $T_1$ and $T_2$ of the two bodies and their separation $L$, the total heat flow is
\begin{equation}
Q = \int_0^{\infty} S_{\omega} d\omega,
\label{Qt}
\end{equation}
where $\omega$ is the frequency of the electromagnetic modes between the perforated slabs, and $S_{\omega}$ is the spectral heat flux:
\begin{align}\label{eq:Sw}
S_{\omega}(\omega,L,T_1,T_2) &= \left[ \Theta\left(\omega,T_1\right) - \Theta\left(\omega,T_2\right) \right] \\
&\times \int_0^{2\pi} \int_0^{\infty} \frac{d\beta \, \beta \, d\phi}{(2\pi)^2}
\left[ \tau^{\mathrm{prop}}\left(\omega,\kappa_0,L\right) +
       \tau^{\mathrm{evan}}\left(\omega,\kappa_0,L\right) \right], \nonumber
\end{align}
where $\beta$ is the parallel wavevector component, $\phi$ is the azimuthal angle, and 
$k_z = \sqrt{(\omega/c)^2 - \beta^2}$ is the normal component of the wavevector. The Planck distribution is 
\[
\Theta(\omega,T) = \frac{\hbar \omega}{\exp(\hbar \omega/k_B T) - 1},
\]
where $k_B$ is Boltzmann's constant and $T$ is the temperature.

The energy transmission coefficients $\tau^{\mathrm{prop}}$ (propagating) and $\tau^{\mathrm{evan}}$ (evanescent) can be expressed in terms of the reflection matrices $\mathbf{R}_{1,2}$ of the two slabs:
\begin{equation}
\tau_j^{\mathrm{prop}} = \mathrm{Tr}\left[ (\mathbf{I} - \mathbf{R}_2^{\ast} \mathbf{R}_2)
\, \mathbf{D} \, (\mathbf{I} - \mathbf{R}_1 \mathbf{R}_1^{\ast}) \, \mathbf{D}^{\ast} \right],
\label{taup}
\end{equation}
\begin{equation}
\tau_j^{\mathrm{evan}} = \mathrm{Tr}\left[ (\mathbf{R}_2^{\ast} - \mathbf{R}_2)
\, \mathbf{D} \, (\mathbf{R}_1 - \mathbf{R}_1^{\ast}) \, \mathbf{D}^{\ast} \right] e^{-2|k_z|L},
\label{taue}
\end{equation}
where the reflection matrix for slab $i = 1, 2$ is
\begin{equation}
\mathbf{R}_{1,2} =
\begin{pmatrix}
r_{pp} & r_{ps} \\
r_{sp} & r_{ss}
\end{pmatrix},
\label{refm}
\end{equation}
with $p$ and $s$ denoting the polarization of light. Each element $r_{\alpha \beta}$ represents the reflection coefficient from incident polarization $\alpha$ to reflected polarization $\beta$. The matrix 
\begin{equation}
\mathbf{D} = \left( \mathbf{I} - \mathbf{R}_1 \mathbf{R}_2 e^{2ik_z L} \right)^{-1},
\label{Dfp}
\end{equation}
describes the Fabry–Pérot resonances between the plates, and $\mathbf{I} $ is a $2\times2$ identity matrix.

The validity of the effective medium approximation (EMA) in the near-field regime requires that the characteristic inhomogeneity length scales remain much smaller than the relevant electromagnetic decay lengths. In our geometry, the two key scales are the average separation between the holes $\langle R \rangle$ and the characteristic  size of the hole $a$, while the dominant lateral wavevectors contributing to the heat-transfer spectrum are of order $\beta \sim 1/L$, with $L$ denoting the vacuum gap thickness. 
A practical criterion for the applicability of EMA is therefore
\begin{equation}
\qquad
  \beta <\frac{1}{\langle R \rangle},
\end{equation}
ensuring that the evanescent waves responsible for most of the transfer cannot resolve the fine structure of the perforations.  The effective wavelength inside the medium is $ \lambda_\parallel$, 
\begin{equation}
  \lambda_\parallel = \frac{2\pi}{\beta},
\end{equation}
should satisfy $\lambda_\parallel > a$, implying that 
\begin{equation}
 a\beta<2\pi.
 \end{equation}

 In addition, the condition
\begin{equation}
  L \gtrsim \frac{\langle R \rangle}{\pi}
\end{equation}
has been suggested in previous homogenization studies \cite{biehs2011nanoscale} to prevent strong coupling to higher diffraction orders.  
 In the calculations presented in the next section we have to consider first that the integration over frequency in Eq. (\ref{Qt}) can be integrated up to an upper cut-off frequency since the spectral function $S(\omega)$. For the integration over the wavevector $\beta$, the upper bound of integration is limited to $\beta<2\pi/a$. 
This requirement implies that, when evaluating the integral over 
$\beta$  in Eq. (9), the upper limit of integration must lie within the regime where homogenization is reliable, while also capturing the $\beta$ -range that contributes most of the heat flux.  In the Supplementary section, we present validity tests for the EMA for the parameters  used in this work. 

Another point of discussion is whether the array of holes should be treated as periodic or randomly distributed. Within the EMA framework, both cases reduce to the same description provided the filling fraction is unchanged. However, this neglects resonant phenomena such as extraordinary optical transmission (EOT), which may arise in periodic perforated plates, particularly in the Reststrahlen band of SiC due to coupling with surface phonon polaritons. In disordered (random) hole distributions, these collective lattice resonances are largely suppressed, leading to a weaker manifestation of EOT \cite{ebbesen1998extraordinary,braun2011optical,garcia2010light}.

\section{SiC and Near Field Heat Transfer} 
As a first example, we consider a perforated SiC plate. The cylindrical inclusions are assumed to be empty, so that $\epsilon_i=1$, while the dielectric response of SiC is described by
\begin{equation}
\epsilon_h=\epsilon_{\infty}\left ( 1+\frac{\omega_L^2-\omega^2-i\gamma \omega}{\omega_T^2-\omega^2-i\gamma \omega}\right),
\end{equation}
where $\epsilon_{\infty}=6.7$, longitudinal phonon frequency $\omega_L=1.827\times10^{14}$ rad$\cdot$s$^{-1}$, transverse phonon frequency $\omega_T=1.495\times10^{14}$ rad$\cdot$s$^{-1}$, and damping constant $\gamma=0.9\times10^{12}$ rad$\cdot$s$^{-1}$.

We consider two parallel slabs separated by a vacuum (or dielectric) gap. Each slab is perforated by an ensemble of parallel cylinders oriented perpendicular to the slab surfaces.  The cylinders may have a circular cross section of radius 
$r$ or an elliptical cross section with semi-axes $a$ and $b$.
Their centers are randomly distributed within the slab, yielding a total inclusion fraction equal to the volume fraction $f$. In order to isolate the role of anisotropy, all comparisons between circular
and elliptical perforations are performed at fixed filling fraction $f$ and
slab thickness $d$. To ensure that the same filling fraction is maintained when changing the shape of the cross section, we choose the parameters of the inclusions so that the area of the cylinders cross section remains constant as well as the assumed thickness of the slab. 
The orientation of the ellipse introduces tunable in-plane anisotropy: the major axis forms an angle $\phi$ with respect to the $y$-axis, such that for $\phi = 0$ it aligns with the $y$-axis, while for $\phi=90^o$ it aligns with the $x$-axis.

\begin{figure}[ht]
\centering
  \includegraphics[width=0.75\textwidth]{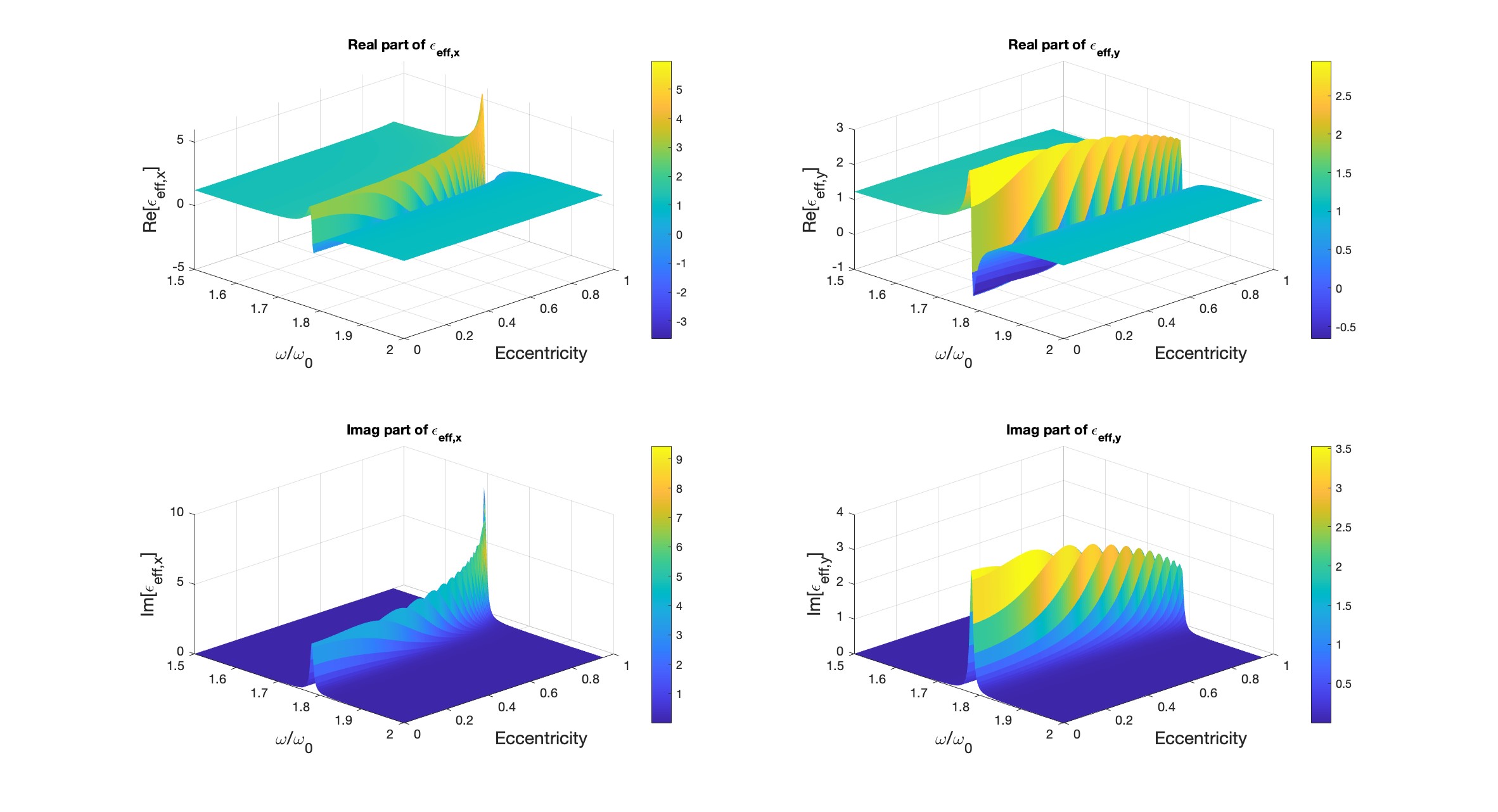}
  \caption{The real (top) and imaginary parts (bottom) of the effective dielectric function of a slab along the $x$ and $y$ axes,  with cylindrical  perforation of different eccentricity. The slab is made of SiC and the holes are empty (air). The frequency is normalized to $\omega_0=10^{14}$ rad$\cdot$s$^{-1}$. The perforations are assumed to be randomly distributed. }
  \label{Figepsr}
 \end{figure}

First we present the effective dielectric function as a function of the normalized frequency $\omega/\omega_0$, where $\omega_0=10^{14}$ rad$\cdot$s$^{-1}$, and as a function of eccentricity.  In Figure (\ref{Figepsr}), we show the real and imaginary parts of the dielectric functions,  along the $x-$axis and $y-$axis. The color bars indicate the value of the dielectric function and each has different size. Since we are considering SiC, most of the changes occur in the frequency region within the Reststrahlen band.  
 
\begin{figure}[ht]
\centering
  \includegraphics[width=0.6\textwidth]{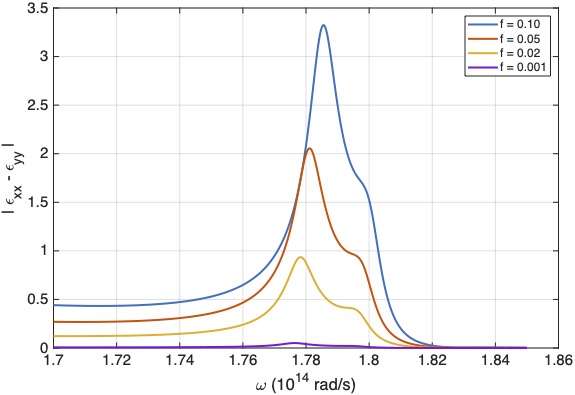}
  \caption{ Anisotropy of the dielectric function $\Delta_{\epsilon}=|\epsilon_{xx}-\epsilon_{yy}|$ as a function of frequency for different filling fractions and at a fixed eccentricity of $e=0.67$.  As the value of $f$ decreases the anisotropy also decreases.  }
  \label{Figepsanis}
 \end{figure}

 The effective dielectric function  depends in the filling fraction and the eccentricity of the elliptical cylinders.  For $f=0$ the homogeneous isotropic host dielectric function is recovered. Similarly, for any value of the filling fraction $f$, the isotropic case is recovered  when we have circular perforations in the host materials. For a fixed value of $f$, the eccentricity will determined the degree of anisotropy of the dielectric function. To quantify this, we define the difference $\Delta_{\epsilon}=|\epsilon_{xx}-\epsilon_{yy}|$ as the anisotropy factor. In Fig.(\ref{Figepsanis}) we plot the anisotropy factor as a function of frequency, for different values of $f$. The eccentricity for this plot is $e=0.67$ and the ellipsoidal cross sections are oriented along the $x$-axis. As expected, as the value of $f$ decreases the anisotropy factor becomes smaller.   

As an aid we plotted the maxima of $\Delta_{\epsilon}$ as a function of eccentricity and filling fraction. This is shown in figure S2 of the Supplementary material. The regions where the anisotropy factor is bigger can be determined for specific values of eccentricity and filling fraction.  

\begin{figure}[ht]
\centering
  \includegraphics[width=0.6\textwidth]{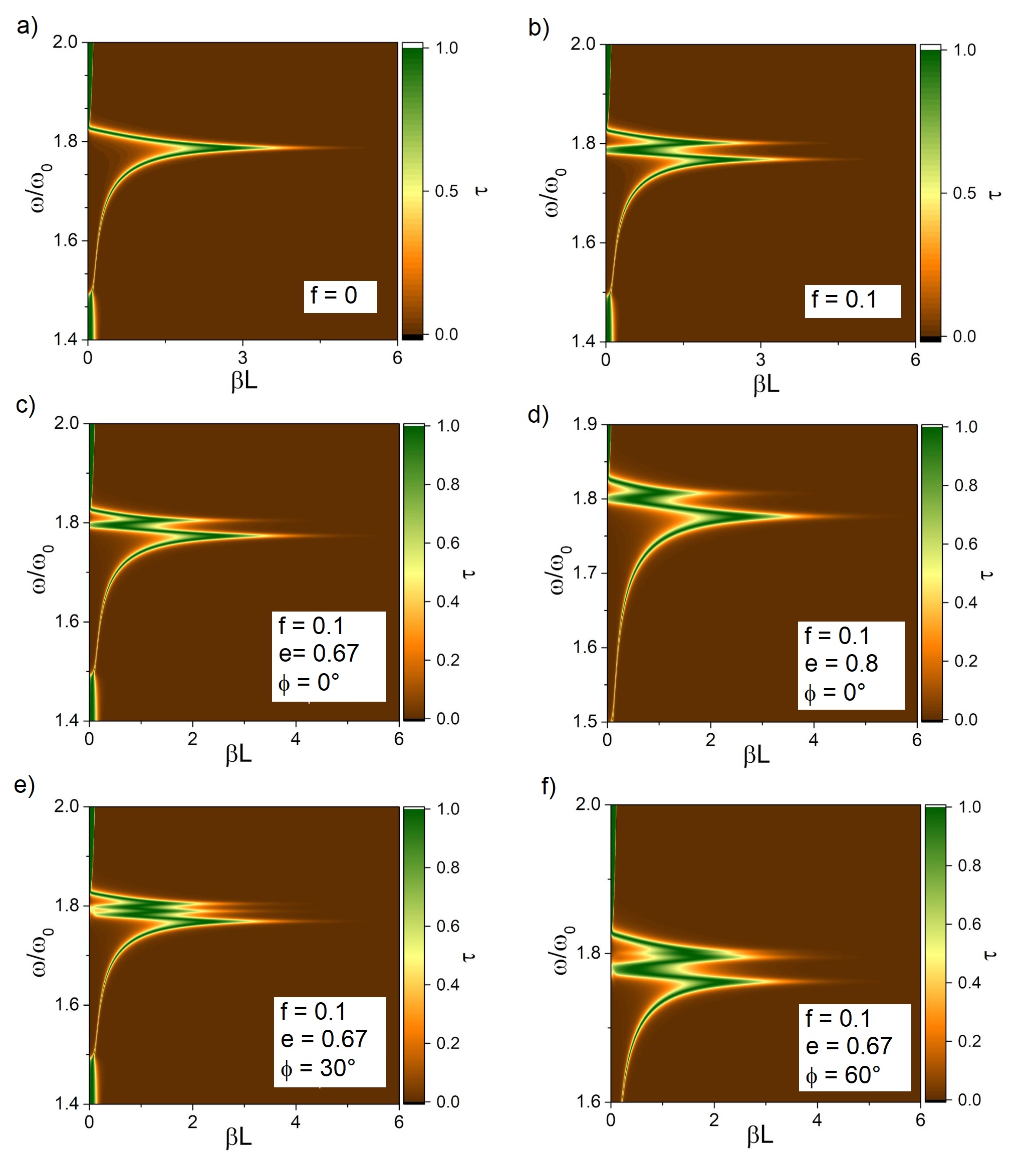}
  \caption{ Transmission coefficient for two parallel slabs with cylinders of circular cross section. (a) SiC with no holes $f=0$ and (b) with circular cylinders and volume fraction $f=0.1$. The perforations are assumed to be randomly distributed.  }
  \label{Figtaus}
 \end{figure}

To study the NFRHT, we consider first the 
 case of circular cylinders that was previously analyzed by S.-A. Biehs \textit{et al.} (2011) \cite{biehs2011nanoscale}. For reference,  we reproduce their calculations here. In Figure~\ref{Figtaus} the transmission factors Eqs.~(\ref{taup},\ref{taue}) are plotted. Fig.~(\ref{Figtaus})(a) is just for the SiC slab with no perforations and (b) when the plate has a volume fraction of $f=0.1$.  The splitting of the mode becomes apparent. Keeping the filling fraction at $f=0.1$ we change the eccentricity of the cylinders to $e=0.67$ (c) and $e=0.8$ (d). The rotation angle is $\phi=0$. We observe changes in the transmission around $\omega/\omega_0=1.8$ 
and at low frequencies and low values of wavevector. 
Also, in Fig. (\ref{Figtaus}) (e,f) the energy transmission is plotted for a fixed value of the filling fraction ($f=0.1$), same eccentricity ($e=0.67$) but for different angles. The transmission energy in (c, e and f) changes and at an angle of $\phi=30^{o}$ three modes become clear. 

\begin{figure}[t!]
\centering
  \includegraphics[width=0.6\textwidth]{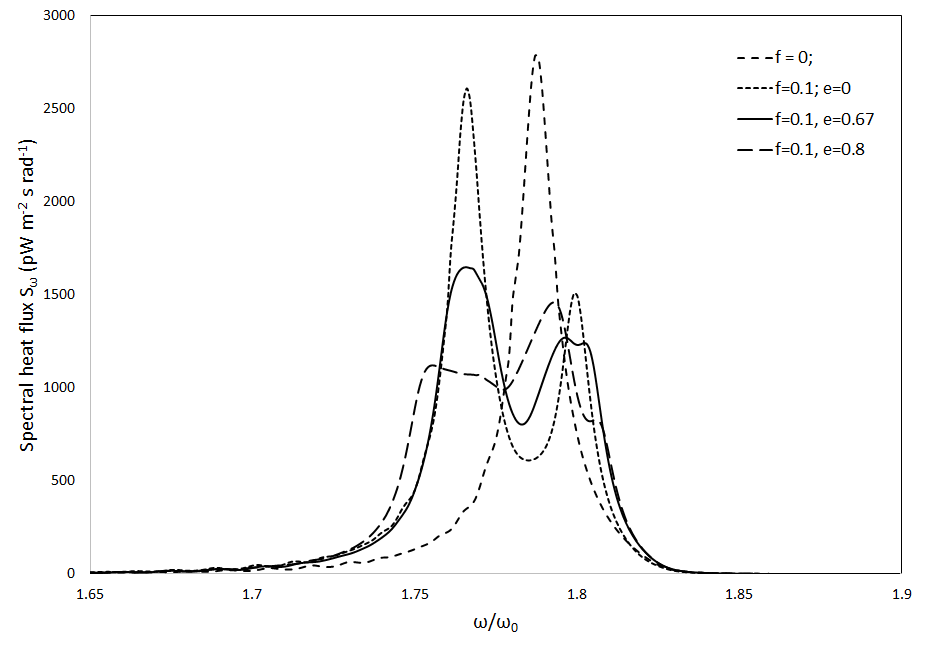}
  \caption{Spectral heat flux $S_{\omega}$ for SiC slabs ($f=0$) and for perforated slabs with the same filling fraction of $f=0.1$ but different cross section circular (dashed line), elliptical with $e=0.67$ solid line and $e=0.8$ long-dashed line. The separation between the plates is $L=100$ nm. The perforations are assumed to be randomly distributed. }
  \label{Figspec}
 \end{figure}

 To further study the effect of the eccentricity, we present in Fig.~(\ref{Figspec}) the spectral heat flux $S_{\omega}$ as a function of the normalized frequency $\omega/\omega_0$. The plates are at a separation of $L=100$ nm. As a reference we plot $S_{\omega}$ only for the SiC plates ($f=0$) and the other curves correspond to the same filling fraction for $f=0.1$ but with circular holes ($e=0$ dashed line), cylinders of elliptical cross section ($e=0.67$ solid line) and ($e=0.8$ long-dashed line). As the eccentricity increases we observe the two main peaks present for circular cylinders decrease in amplitude, but start increasing in width.
  It is important to mention that when varying the value of the eccentricity, the semi-axes $a$ and $b$ are chosen so that the cross section $\pi ab$ remains constant and equal to the area of the circular cylinder to maintain the same value of $f$. Therefore, variations in the results arise solely from shape-induced anisotropy rather than changes in inclusion volume or slab thickness.

Keeping the eccentricity fixed the spectral heat flux $S_{\omega}$ is plotted as a function of frequency for different filling fractions $f$.  Three cases are shown $f=0$ (homogeneous slab of SiC), $f=0.05$ and $f=0.1$ for an eccentricity of $e=0.67$.  Introducing a finite concentration of elliptical holes increases the 
in-plane anisotropy through distinct depolarization factors 
$L_x \neq L_y$. As the filling fraction increases, both peaks 
decrease in amplitude, broaden, and undergo slight frequency shifts, 
reflecting the modified effective permittivities $\varepsilon_{xx}$ 
and $\varepsilon_{yy}$ of the anisotropic composite. 

\begin{figure}[t!]
\centering
  \includegraphics[width=0.6\textwidth]{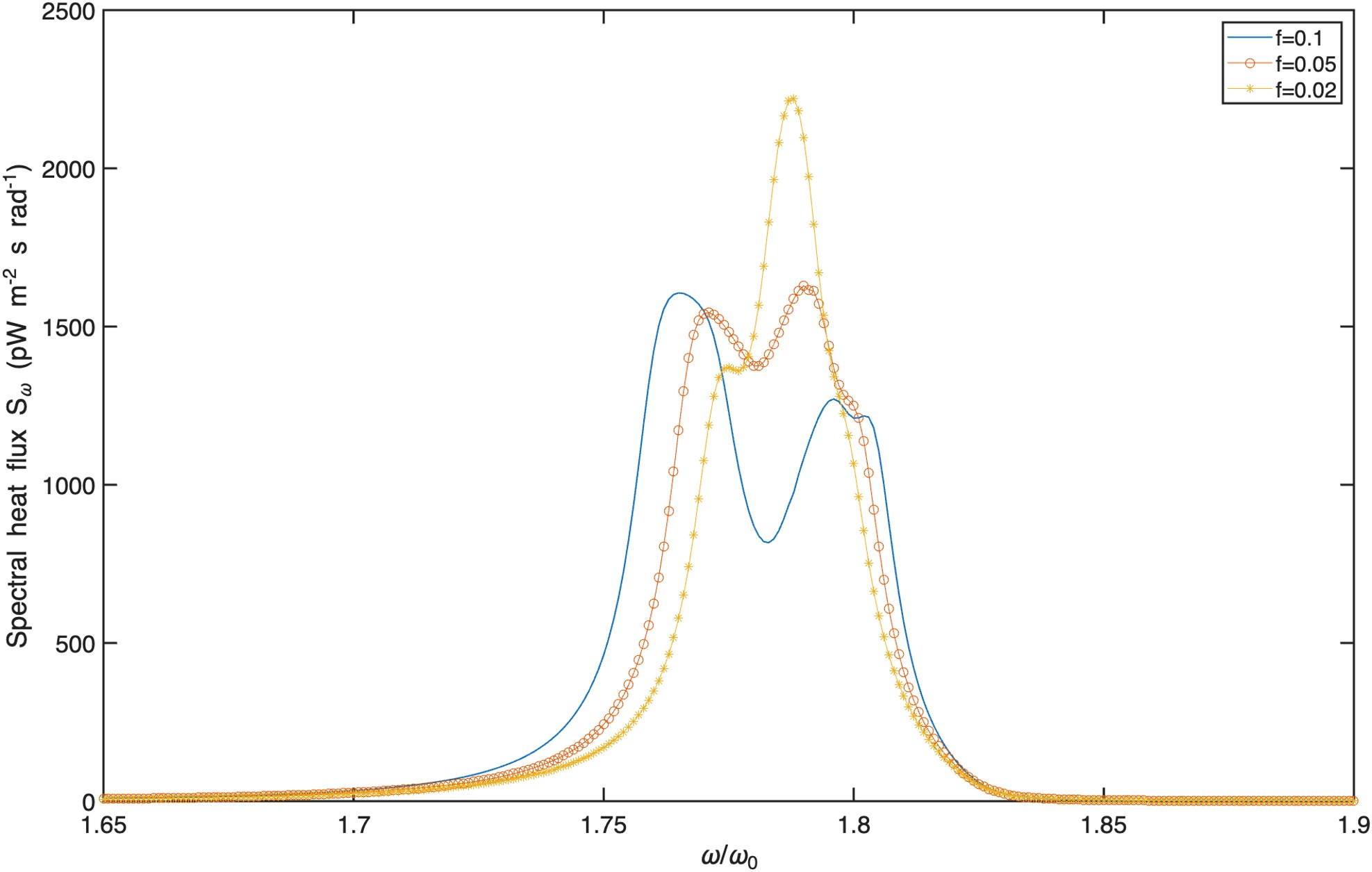}
  \caption{Spectral heat flux $S_{\omega}$ as a function of the normalized 
frequency $\omega/\omega_{0}$ for bulk SiC ($f=0$) and for composite 
SiC slabs perforated with elliptical cylindrical holes of fixed 
eccentricity $e=0.67$ and filling fractions $f=0.05$ and $f=0.1$. }
  \label{Figswf}
 \end{figure}

\begin{figure}[ht]
\centering
  \includegraphics[width=0.6\textwidth]{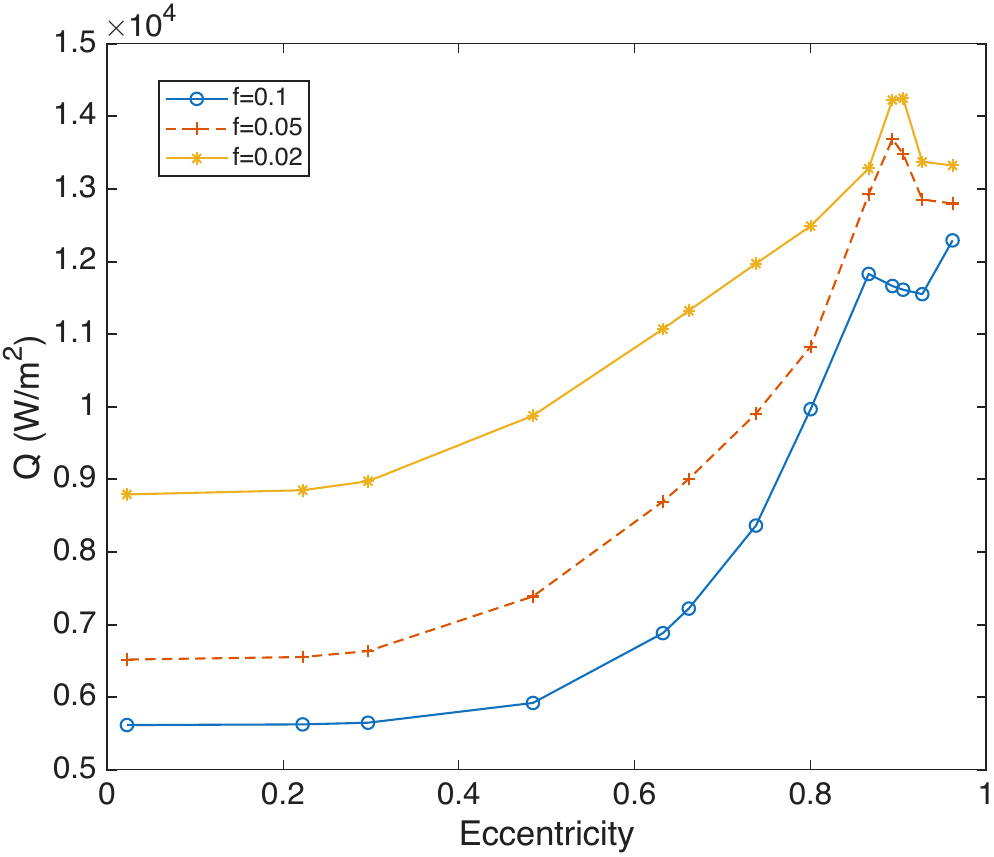}
  \caption{Total heat flux as a function of eccentricity of the cylinders when the slabs have a constant separation of $L=100$ nm. The different curves are for s filling fractions of $f=0.1, 0.05$ and $0.02$. The perforations are assumed to be randomly distributed and oriented along the $x$-axis.   }
  \label{Figtotal}
 \end{figure}

Finally, we calculate the total heat flux keeping the separation constant ($L=100$ nm) and varying the eccentricity of the cross section of the cylinders. The mayor axis is oriented along the $x$-axis.  The filling fraction for the three curves are   $f=0.1,0.05,.02$. As seen in Figure (\ref{Figtotal}), for all values of $f$, the total heat remains almost constant up to $e=0.3$, after which the total heat flux increases monotonically, up to $e=0.83$. The increase of the total near-field heat flux with eccentricity can be attributed to the progressive anisotropy introduced by elliptical perforations. As  discuss earlier, as $e$ increases, the in-plane permittivity tensor becomes increasingly biaxial, which broadens the Reststrahlen resonance and opens additional channels for tunneling of surface phonon polaritons across the vacuum gap. This enhances the density of contributing modes and results in the observed rise of the total flux. 

 In the region $0.83<e<1$ the observed behavior is different, due to two main reasons. 
First, the depolarization factors  saturate in the extreme-ellipse limit, this is, they approach their asymptotic values of $L^{(x)}\rightarrow 0$ and $L^{(y)}\rightarrow 1$.  Thus, further increases in aspect ratio no longer generate proportionally larger anisotropy. Second, excessive elongation reduces the effective filling fraction of the polaritonic material along one axis, thereby diminishing the coupling strength of surface resonances for certain polarizations. The combined effect is a reduction in the overlap between the evanescent modes supported by the two slabs, which manifests as the dip in the total heat flux. In other words, there is an optimal eccentricity where anisotropy-induced broadening maximizes the modal overlap; beyond this point the effective medium becomes too dilute along one direction, and the coupling efficiency decreases.

\begin{figure}[ht]
\centering
  \includegraphics[width=0.6\textwidth]{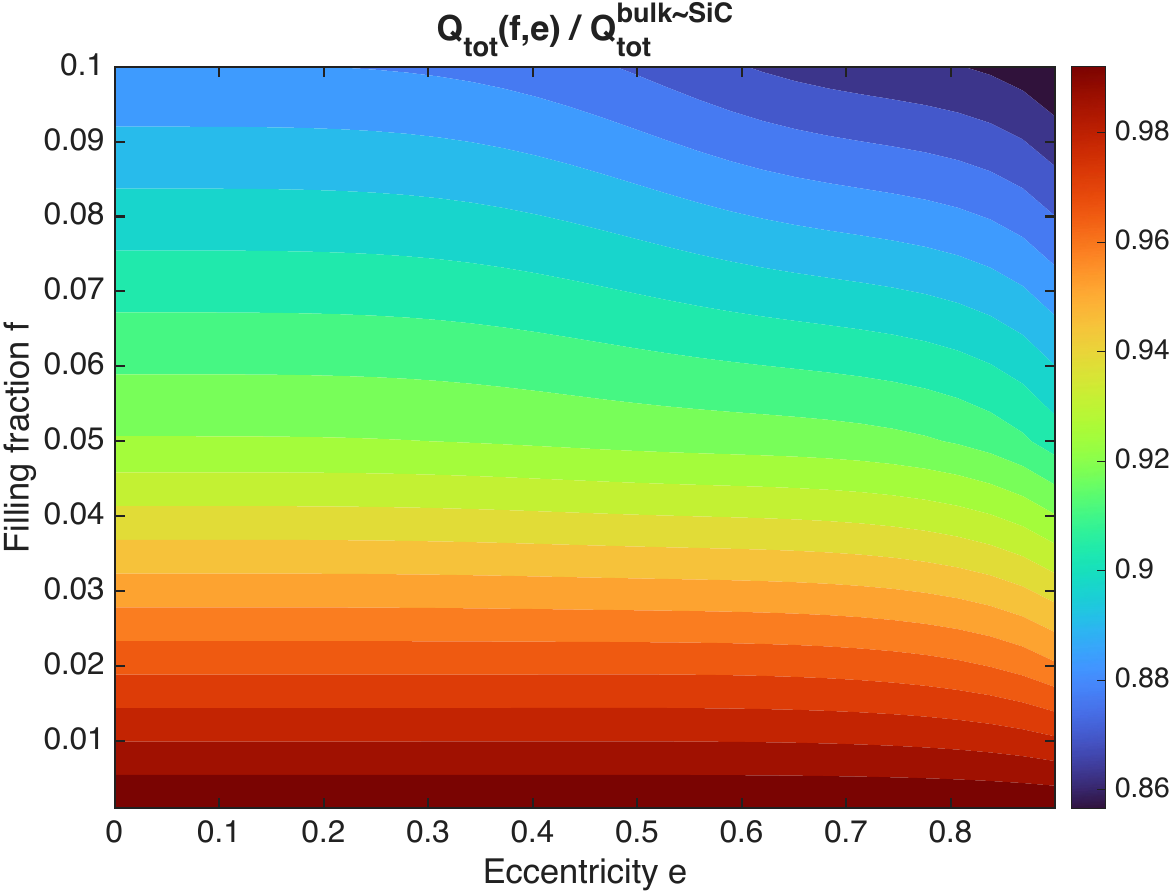}
  \caption{Map of the total heat transfer normalized to the value of bulk SiC as a function of eccentricity and filling fraction. The separation between the plates is $L=100$ nm. }
  \label{Qfemap}
 \end{figure}

A more general picture of the behavior of the total heat flux as a function of eccentricity and filling fraction is presented in Fig.(\ref{Qfemap}), where the total heat flux normalized to the bulk value of SiC is shown as a function of filling fraction and eccentricity. In this case the separations between the plates is $L=100$ nm.  

\begin{figure}[ht]
\centering
  \includegraphics[width=0.65\textwidth]{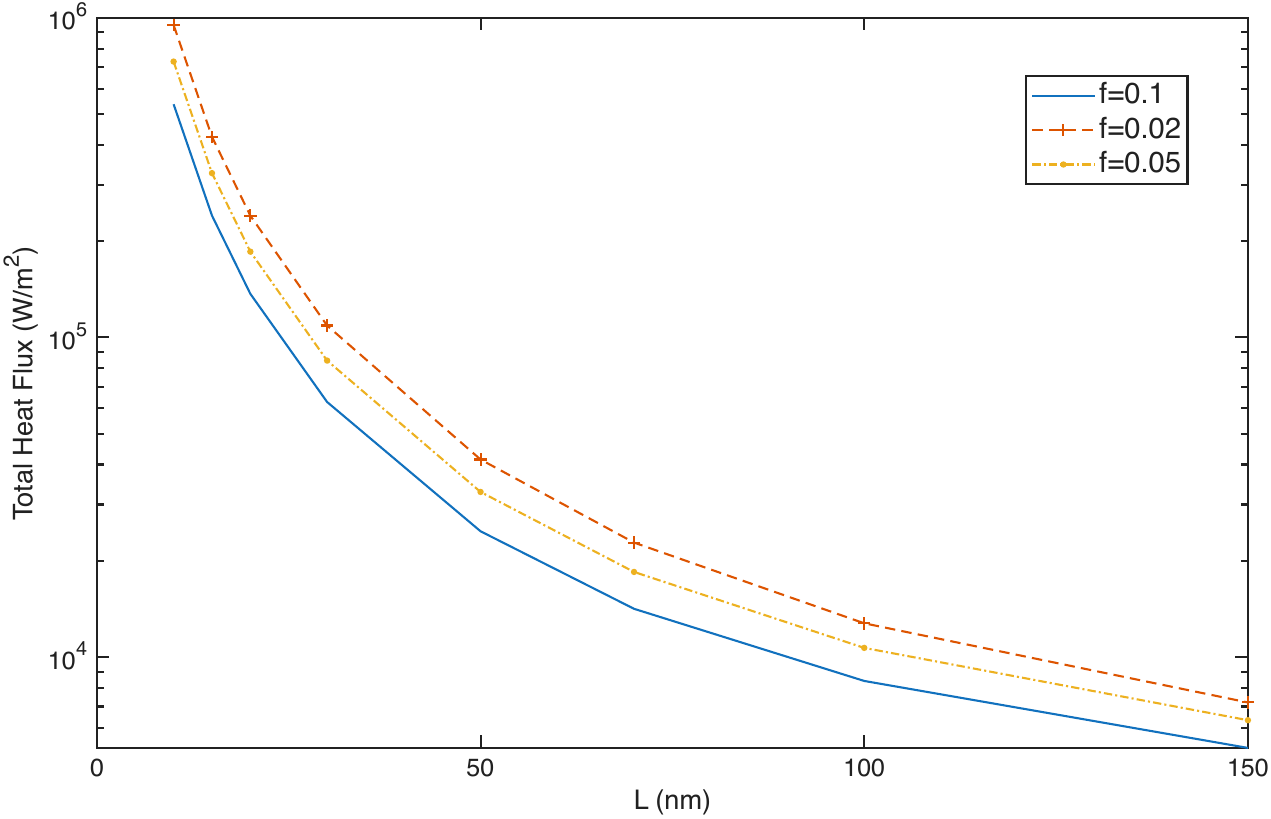}
  \caption{Map of the total heat transfer normalized to the value of bulk SiC as a function of eccentricity and filling fraction. The separation between the plates is $L=100$ nm. }
  \label{QfL}
 \end{figure}

Finally, and for completeness, we study the dependence with plate separation of the total heat transfer. In this case, the dependence with separation is not expected to change since by introducing the perforations we are only varying the effective dielectric function.   
The  plate separation enters only in the exponential factors in Eq.(\ref{taue}) and (\ref{Dfp}). To show this, in Fig.(\ref{QfL}) we show the total heat flux as a function of separation for different filling fractions keeping the eccentricity fixed.  The role of the filling fraction is to change the magnitude of the total heat only.  It is important to notice that in order to calculate the total heat as a function of separation, the conditions for the validity of EMA discussed earlier have to be satisfied at each separation $L$. In an actual experimental situation, since the size of the inclusions is fixed, this will limit the range of separations that can be studied. 

\section{Conclusions}
We have investigated theoretically the near-field radiative heat transfer between parallel slabs perforated with elliptical holes, modeled within an effective medium formalism. By tuning the eccentricity of the perforations, the system transitions from isotropic to biaxial response, giving rise to marked modifications of the transmission spectrum and the total heat flux. Our results show that eccentricity-induced anisotropy broadens the Reststrahlen-band resonances of SiC, thereby enhancing the overall heat flux compared to circular perforations.
Also, the total flux exhibits a non-monotonic dependence on eccentricity: it increases up to an optimal aspect ratio where anisotropy maximally enhances coupling, and decreases for larger eccentricities as effective dilution along one axis reduces the modal overlap.  The angular dependence introduced by elliptical holes provides a practical route to directional control of near-field heat transfer, suggesting potential for anisotropic thermal management at the nanoscale.

 The theoretical framework employed here, Maxwell-Garnett effective 
medium theory combined with fluctuational electrodynamics, is scale invariant 
in the quasi-static limit and therefore applies directly to experimentally 
realistic pore sizes. In practice, feature diameters in the range of 
$50$-$60~\mathrm{nm}$ are readily achievable in SiC using electron-beam 
lithography (EBL) \cite{rooks1997electron}, nanoimprint lithography (NIL) \cite{schift2017nanoimprint}, or optimized focused-ion-beam 
(FIB) milling \cite{orloff2017handbook}. For such dimensions the condition for homogenizability, 
$\beta a \ll 1$, remains satisfied at gap distances of $L \sim 
100~\mathrm{nm}$ and higher, since the dominant evanescent parallel wavevectors lie in the 
range $\beta \sim 5$-$8\,k_0$. Consequently, the effective 
dielectric response depends primarily on the filling fraction and eccentricity 
rather than on the absolute pore diameter, and the anisotropy-induced 
modifications of the near-field heat transfer predicted here remain valid for 
pore sizes in the $50$-$60~\mathrm{nm}$ regime accessible to fabrication. Thus, the parameter space explored is not only theoretically meaningful but also aligned with practically realizable structures. 

Overall, our  study demonstrates that controlled anisotropy, engineered through elliptical perforations, can be used  for tailoring near-field thermal radiation, opening avenues toward directional thermal devices and anisotropic thermal metamaterials.

\section{Supplementary Material}   In the Supplementary Material the derivation of the effective medium approximation for cylinders of elliptical cross section is derived, as well as the expressions for the depolarization factors. An anisotropy map for different filling fraction and eccentricities is included. Finally, the convergence tests and validity tests for using effective medium approximations are presented.

%

\end{document}